\documentclass[final,3p,times]{elsarticle}

\usepackage{amsmath}
\usepackage{hyperref}
\usepackage{natbib}
\usepackage{xcolor}
\usepackage{amssymb}
\biboptions{sort&compress}

\newcommand{\ba}{\begin{eqnarray}}
\newcommand{\ea}[1]{\label{#1} \end{eqnarray} }

\begin{document}
%\linenumbers
\begin{frontmatter}


%% \ead{zneda$@$phys.ubbcluj.ro}[label1]

\title{\bf \Large Jackpot statistics, a physicist's approach}


\author[label1]{Istv\'an Gere \corref{cor1}},
\author[label1]{Szabolcs Kelemen} 
\author[label1]{Zolt\'an N\'eda}
\author[label2,label3]{and Tam\'as S. Bir\'o}

\address[label1]{Physics Department, Babe\c{s}-Bolyai University, Cluj-Napoca, RO-400347, Romania}
\address[label2]{HUN-REN Wigner Research Centre for Physics, H-1525 Budapest, P.O.Box 49, Hungary}
\address[label3]{ Complexity Science Hub, Vienna, Austria}

\cortext[cor1]{Corresponding author: istvan.gere$@$ubbcluj.ro}

\begin{abstract}
At first glance lottery is a form of gambling, a game in which
the chances of winning are extremely small.
But upon a deeper look, considering that the Jackpot prize of lotteries is a
result of the active participation of millions of players, we come to the
conclusion that the interaction of the simple rules with the high number of
players creates an emergent complex system.
Such a system is characterized by its time-series that presents some
interesting properties. Given the inherent stochastic nature of this game,
it can be described within a mean-field type approach, such as
the one implemented in the Local Growth and Global Reset (LGGR) model.
We argue that the Jackpot time-series behaves ergodic for
six lotteries with diverse formats and player pools.
Specifying this consideration in the framework of the LGGR model,
we model the lotteries with growth rates confirmed by the time-series.
The reset rate is deduced mathematically and confirmed by data.
Given these parameters, we calculate the probability density of the Jackpot prizes,
that fits well the \textcolor{black}{empirically} observed ones.
%We further discuss the way the model parameters relate to the dynamics of the lottery.
We propose to use a single $w$ parameter, as the product of the player pools found under
the jurisdiction of the lottery and the chance that a single lottery ticket wins.
\end{abstract}



\begin{keyword}
growth and reset process, master equation, stationary distributions, transient dynamics



\end{keyword}

\end{frontmatter}
 %\linenumbers

%\section{}
%\label{}



%%       \citet{<label>} ==> Jones et al. [21]
%%       \citep{<label>} ==> [21]
%%



\section{Introduction}
Making decisions based on chance has always been a practical solution applied by people
in numerous territories of life.
Such a practice in ancient times was the basis of divination methods, such as the I Ching~\cite{iching}, but they also came \textcolor{black}{in} handy to modern physicists in the form of Monte-Carlo simulations. It may be used to settle disputes, assign land, or work~\cite{lotto_hist}.
In this sense, lotteries constitute a type of gambling where the fundamental mechanism
involves the random choice of a winner from a pool of players.
In this paper we look at pseudo-active lotteries, where players buy a chance to win
a Jackpot prize by selecting a series of numbers on a ticket.
When no winner occurs, the Jackpot prize (the prize pool) grows incrementally,
with a given portion of the weekly sales being added to it.
Winners are selected through a periodic random number draw, which is scheduled at
regular intervals, weekly or bi-weekly in general.
If a ticket (or tickets) happens to contain the identical numbers as those drawn,
the winner will receive the prize pool (in case of multiple winners they share the prize),
whereas in the absence of a winner, the prize pool persists in growing through each
successive draw until a winner is finally declared~\cite{lotto_chance}.
One of the most popular lotteries is the '6/49' Lotto, in which participants choose
six numbers from a range of 1 to 49, explaining its '6/49' designation~\cite{lotto_gen}.
Depending on the specific lottery rules, participants who match fewer numbers than
what is required to win the Jackpot (but still more than a minimum threshold) may be
eligible for smaller prizes. These smaller prizes can be funded from either a portion
of the Jackpot prize pool or from a dedicated fraction of the weekly sales.
Jackpot lotteries, such as the Powerball in the United States, guarantee a weekly
increase of the prize value until someone wins~\cite{Powerball_growth,Powerball_rules}.

Throughout history and in contemporary times, lotteries have typically been administered
by states, providing a consistent source of revenue for state budgets, with usually only
a fraction (typically $50\%$) of the sales contributing to the prize
pool~\cite{Powerball_rules,Megamill_rules}.
The remaining portion of the sales is allocated to cover the administrative expenses of
the lottery, which are then legally taxed. Additionally, when a winner is chosen,
the Jackpot prize is also subject to taxation~\cite{lotto_hist}.
The extremely slim chances of winning the lottery
(in the case of $6 / 49$ lotto there are $C(49,6) = 13983816$ distinct combinations
for selecting $6$ numbers from $49$, meaning that the chance of winning with a single
ticket is roughly 1 to 14 million)\cite{lotto_gen,lotto_chance} allows the jackpot
to grow after consecutive plays, attracting even more players~\cite{growth_and_addiction}.
This high Jackpot, coupled with the extremely small chances to win,  fascinates people,
making lotteries a popular and addictive game~\cite{growth_and_addiction}.

To provide an example, the North American Association of State and Provincial Lotteries
reported that United States Lotteries achieved total sales of $\$98,519,173,817$ in the
fiscal year of 2021~\cite{american_lottos}.
Lotteries, as we can observe, can be approached via their mathematical
aspects~\cite{lotto_chance} (combinatorics, chance of winning, etc.),
psychological aspects (why do people play)~\cite{lotto_gen},
economical aspects (optimizing their rules in favor of state revenue)~\cite{lotto_econ},
and from a statistical physics viewpoint, as we argue in this article.

Approaching lotteries as complex systems, by focusing on the emergent dynamics resulted
from the combination of a set of rules and the decisions of millions of players,
and analyzing them using the methodology of statistical physics,
lends novelty and an intriguing nature to the present study.

\begin{figure}[!ht]
    \centering
    \includegraphics[width=\textwidth]{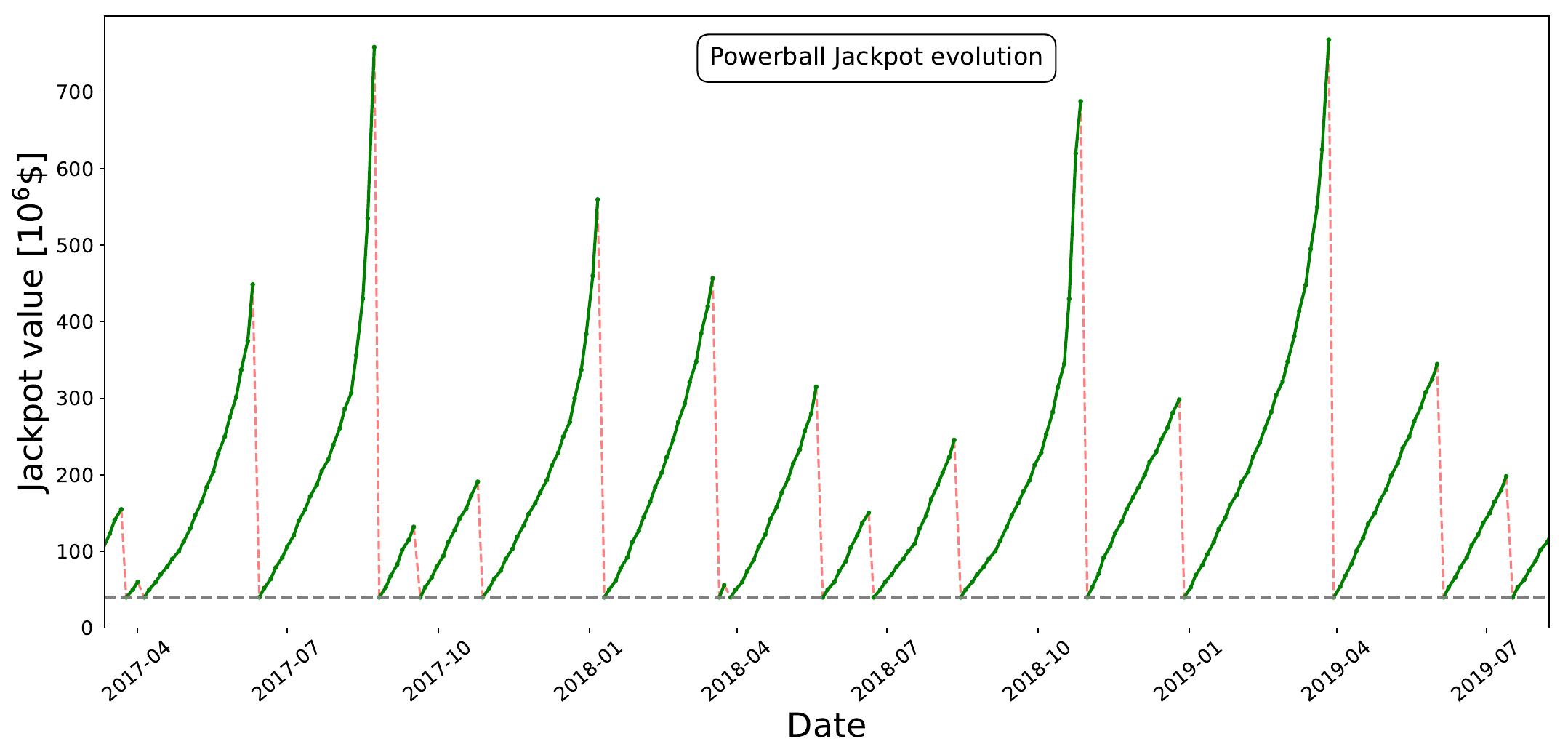}
    \caption{Time series of the Jackpot prize, for the Powerball from 03.2017 to 07.2019.~\cite{canada_dataset}}
    \label{fig1_canda}
\end{figure}

Figure~\ref{fig1_canda} presents a part of the Jackpot value time-series for the Powerball lotto created from freely available Jackpot dataset, collected by players~\cite{canada_dataset}.
Such time series of the Jackpot values summarize the dynamics emerging from the
actions of a huge number of players.
The time evolution of the Jackpot presented in Figure~\ref{fig1_canda} consists of
qualitatively two different types of regions.
First, there are relatively longer regions dominated by gradual growth of the Jackpot value,
due to the funds accumulated by the sold tickets when no winner tickets were sold
(green segments of the time series).
The regions characterized by a monotonic growth of the Jackpot are augmented by
intervening and prompt resetting events of the Jackpot to a default value,
when the Jackpot is won (red dashed sections of the time series).
The processes controlling lottery jackpot values are identical to those serving as
the foundation for the LGGR  (Local Growth and Global Reset)  mean-field type dynamical
model, built on a master equation, which, akin to the previously described,
comprises alternating probabilistic growth and reset processes over time~\cite{LGGR}.
The LGGR approach has recently been applied successfully to demonstrate the underlying dynamics of a decent number of phenomena, ranging from the field of economics through the field of social sciences to biology~\cite{ZNeda1,ZNeda2,ZNeda3,ZNedaFacebook,Kelemen2024}.

\textcolor{black}{
Similar stochastic processes incorporating probabilistic growth and reset dynamics are intensively studied by others as well~\cite{review_stoch}. A special class of such intensively studied processes is the ones characterized by multiplicative growth coupled with reset, resulting in power-law-tailed probability distributions. 
Recent works have studied in-depth analytically and numerically the capabilities of such dynamics to describe phenomena that produce power laws, considering different transition rates~\cite{Black_swan}. Such a dynamics is applied by us also within the framework of the LGGR approach~\cite{LGGR} for modeling income~\cite{ZNeda1} and wealth~\cite{ZNeda2,ZNeda3} distributions in different social systems and the popularity of social media posts and scientific articles~\cite{ZNedaFacebook}. The effect of resetting has also been investigated in replicator dynamics~\cite{entrop}, and a broader review of resetting in stochastic processes is discussed in~\cite{review_stoch}. 
In this article, we apply such processes to model the complex dynamics of lottery Jackpots. Although, the here presented dynamics shows similarities with the ones already explored recently~\cite{review_stoch,Black_swan,entrop}, the reset dynamics characteristic of lotteries is remarkably different. While the focus of our work is partially empirical, based on qualitative observations regarding the real-world system of lotteries, the principles and effects of stochastic resetting described in these works~\cite{review_stoch,Black_swan,entrop} offer a broader context for understanding the consequences of such mechanisms.}  
%The LGGR approach that is the backbone of the theoretical explanations presented in this work comes from a family of models dealing with multiplicative growth coupled with resets, popular in current research as multiple articles~\cite{Black_swan, entrop} and a review\cite{review_stoch} argue for its effective capabilities to describe phenomena that produce power laws. In our previous works, the LGGR model was successfully applied to describe the income distributions in a restricted geographic region \cite{ZNeda1}, the wealth distribution in modern societies \cite{ZNeda2}, and the wealth distribution in an agricultural rural settlement during pre-collectivization, post-communist, and under current free market economic conditions \cite{zneda3}. The model was capable in describing the popularity of social media post and scientific articles \cite{ZNedaFacebook}, and the distribution in biological systems, such as tree sizes in forests \cite{Szabi_erdos_cikk}
%In this article we also consider a dynamics incorporating preferentiality, but
%%%% HERE
This paper is organized as follows:
in Section \ref{sec:Jackpot-ergodicity}
we present the time-series of the Jackpot prize for six different lotteries,
and argue for the ergodicity of these data along the chosen time periods.
In Section \ref{sec:LGGR-general} the LGGR model is presented,
along with the growth and reset rates that are consistent with the dynamics of lotteries.
This section contains the theoretical explanations that lead to the general form of
the stationary probability density function that fits the \textcolor{black}{empirical} results.
Section \ref{sec:discus} is devoted to further discussions on the
chosen rates, also from the perspective of the gathered data.
In the last Section \ref{sec:conclusion} we briefly summarize
the main results of this work.

\section{Jackpot value time-series, and ergodicity
\label{sec:Jackpot-ergodicity}}
\newcommand{\be}{ \begin{equation} }
\newcommand{\ee}[1]{\label{#1} \end{equation} }

To understand and mathematically model the Jackpot prize dynamics,
we analyzed time-series data from various lotteries, each with distinct rule sets
and player
populations~\cite{Powerball_rules,euro_mil_rules,Megamill_rules,texas_rules,Can_lott,Uk_lotto}.
In Table ~\ref{table:lotteries} we present the main statistical properties of the
lotteries investigated in this study for the time periods considered.
The reason for focusing only on limited time windows of lottery histories, instead of studying the available time-series entirely, is that the rules and the territory in which a given lottery is played change from time to time.
Thus, we analyze only time periods with unchanging conditions.
Besides the general information about the lotteries we also list for each of the six lotteries the values given by the product of the Jackpot winning probability
by a single ticket ($P_w$) with the total population of the geographic area where
the given lottery is played ($Pop$).
The expected number of wins is therefore  $w = P_w \times Pop$.
Although this single parameter is not capable of entirely characterizing the lottery by itself, we assume that it conveys valuable information about the lottery.
We further elaborate on the importance of this parameter $w$ in the discussion section.

\begin{figure}[!ht]
    \centering
    \includegraphics[width=\textwidth]{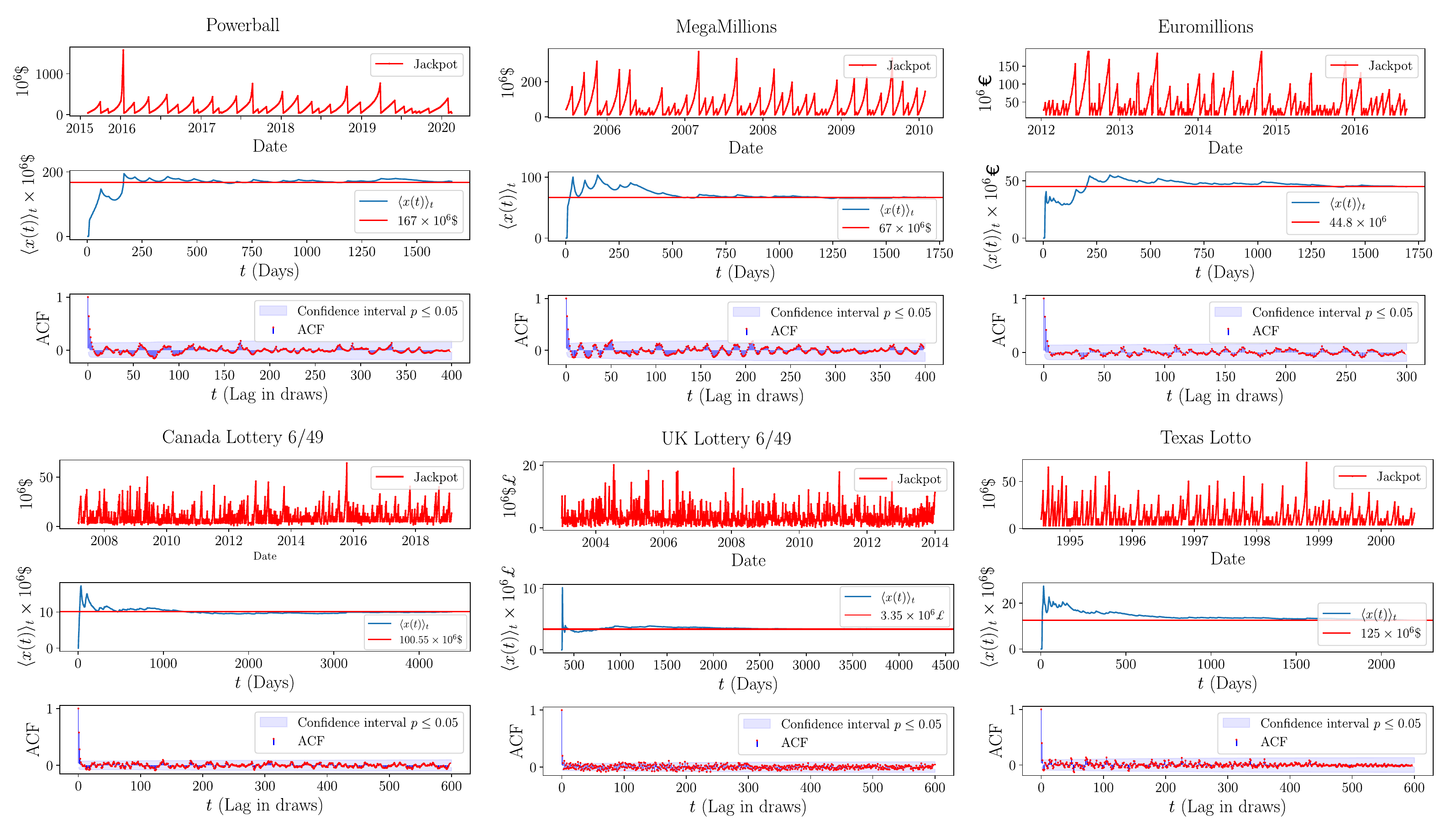}
    \caption{The time series of the lotteries studied (general information provided in
    Table~\ref{table:lotteries}). The rows represent the following: (A.)
	The time series plot of the lotteries, on the time range where the rules and
	the player pools are not changing, (B.)
    the mean of the time series $\langle x(t)\rangle_{t}$ along with the convergence value,
	(C.) the autocorrelation function of the time series $ACF(s)$ until $s$ days
	along with the confidence interval with $p \le 0.05$}
    \label{fig2_lottery_time_series}
\end{figure}

%%%%%%%%% HERE2
The data and the rules of the US lotteries (Powerball, Mega Millions, and Texas Lotto)
were scraped from the \textit{Lottery Report}
web-portal~\cite{Powerball_rules,Powerball_15_17,Powerball_18_22,Powerball_growth,Megamill_01_06,Megamill_06_12,Megamill_rules,texas_92_95,texas_96_98,texas_99_01,texas_rules},
where citizens collect data regarding the lotteries.
In the case of the other lotteries we used the following sources:
the time series for the Uk lotto $6/49$ was obtained from the \textit{BeatLottery}
web-portal~\cite{Uk_lotto} that is a portal similar to \textit{Lottery Report}
for lottery enthusiasts.
The Canada 6/49 lotto data was scraped from the web-portal
\textit{National-Lottery}~\cite{Can_lott}.
The time series of the Euromillions lottery was obtained from the
\textit{Loterieplus} web-portal~\cite{euro_mil_data}, while the rules from the official
Euromillions webpage~\cite{euro_mil_rules}.
In sections A of the combined Figure~\ref{fig2_lottery_time_series} we present the
time-series of the six lotteries collected in Table~\ref{table:lotteries}.

As we have anticipated in the previous section, we plan to apply the LGGR framework
for modeling the Jackpot dynamics that is designed to describe ergodic Markovian processes.
The ergodic property implies, however, that in such systems the time average of a
characteristic quantity is equivalent to the ensemble average of the same quantity.
Accordingly, the stationary probability density over the system's possible states
$\rho(x)$ is calculated as an ensemble average from numerous replicas of the
single-element-system, and is equivalent to the probability density of the
states' occurrences, reflected in the time fractions spent by a single element
of the system in the respective states
(assuming a sufficiently long existence of the system)~\cite{ergodicity}.
In the context of lotteries, the intricate system consists of a single agent
(the Jackpot prize) whose development is probabilistically influenced by the
lottery rules and the players via their ticket purchases.
Thus, before applying the LGGR model, it is necessary to test the ergodic nature
of the Jackpot time-series.

%%% HERE3
To establish the ergodicity of the lottery time-series, we underpin our argument by
demonstrating their stationary characteristics, observed through the mean Jackpot
value's convergence across an extended time window $\langle x(t) \rangle_{t} = c$.
The convergence of the mean Jackpot values in time, for each studied lottery game,
are shown in the B sections of Figure~\ref{fig2_lottery_time_series}. %\textcolor{black}{Due to the relatively short length of the considered periods, the effect of inflation can be neglected. This is also supported by the convergence of the average Jackpot value.}
\textcolor{black}{The convergence of the average Jackpot value also suggests that the effect of inflation on lotteries can be neglected within such short periods.}
Furthermore, we confirm their aperiodic nature by revealing the absence of
a sharp autocorrelation ($ACF(s) = Corr(x(t), x(t+s))$).
As it is shown in the C sections of Figure~\ref{fig2_lottery_time_series},
for relatively long time fractions (after $4$ to $6$ draws) the autocorrelation tends
to a negligible value. For much longer time periods it practically tends to $0$
($lim_{s \to \infty} ACF(s) = 0$).
In general, these two features are the necessary criteria for a Markov chain to be
considered ergodic~\cite{stat_book}.
A similar argumentation was applied to demonstrate the ergodic nature of hydrology
data in ~\cite{rainfall_ergodicity}.
These measures relate to the stability of the statistical moments of the time-series,
denoting stationarity.
In most applications, however, stationary processes without proving their aperiodic nature
are considered to be ergodic~\cite{stat_book}.
In the present study, being these two statistical properties (the mean and the autocorrelation) of the Jackpot time-series proven, we consider the here studied lotteries to be ergodic.

\begin{table}[!ht]
\centering
\fontsize{9pt}{9pt}\selectfont
\begin{tabular}{|c|c|c|c|c|c|}
\hline
Lottery & Studied time-frame & Format & $P_{w}$ & Pop. & w \\ \hline
Powerball~\cite{Powerball_15_17,Powerball_18_22}   &  2015-08-05 to 2020-02-15& $5/69 + 1/26$~\cite{Powerball_rules}  & $\approx 1/(292 \times 10^6)$~\cite{Powerball_rules}    &   $\approx 292.2 \times 10^6 $~\cite{Powerball_rules}   &  $\approx 1$ \\ \hline
     Megamillions~\cite{Megamill_01_06,Megamill_06_12}   & 2005-06-22 to 2010-01-31 & $5/56 + 1/46$~\cite{Megamill_rules} & $\approx1/(258 \times 10^6)$~\cite{Megamill_rules}  & $\approx 152.4 \times 10^6$~\cite{Megamill_06_12}   & $\approx 0.58$  \\ \hline
    Euromillions~\cite{euro_mil_data} &  2012-01-12 to 2016-09-01  &  $5/50 + 2/9$~\cite{euro_mil_rules} & $\approx1/(116 \times 10^6)$~\cite{euro_mil_rules} & $\approx 272 \times10^6 $~\cite{UN_pop} & $\approx 2.34$  \\ \hline
    Canada lotto 6/49~\cite{Can_lott}  & 2007-03-03 to 2019-02-27 & $6/49$   &   $\approx 1/(13 \times 10^6)$   &  $\approx 35.2 \times 10^6$~\cite{UN_pop}    & $\approx 2.7$  \\ \hline
    UK lotto 6/49~\cite{Uk_lotto}    & 1995-01-07 to 2023-09-09  &    $6/49$    &  $\approx 1/(13 \times 10^6)$ &  $\approx 62.5 \times 10^6$~\cite{UN_pop}  &  $\approx 4.8$ \\ \hline
    Texas lotto ~\cite{texas_92_95,texas_96_98,texas_99_01}    & 1994-07-14 to 2000-07-18 &   $6/50$~\cite{texas_rules}  &  $ \approx 1/(17 \times 10^6)$       &      $\approx 18.9\times 10^6$~\cite{texas_pop}    & $\approx 1.12$  \\ \hline
\end{tabular}
\caption{The lotteries studied in this article, along the time frame where the rules are
	consistent. We present the format of the lottery, the chance of winning with a
	single ticket, the possible maximal player population, and the lottery parameter $w$.}
\label{table:lotteries}
\end{table}

\section{The LGGR approach}
\label{sec:LGGR-general}

In order to describe the statistics of Jackpot values, we utilize the LGGR
(Local Growth and Global Reset) model~\cite{LGGR}.
This model accounts for a process with a unidirectional probabilistic growth
and an augmented reset, the same processes that are driving the dynamics of the Jackpot value.
Instead of using an ensemble of elements that all evolve according to
such a dynamics, here we consider a long-term evolution of the Jackpot value,
where the dynamics consists of multiple regions with probabilistic growth concluded
by a reset process, as described in the introduction.

We assume that we have a Markov process characterized by a stationary stochastic matrix.
The stationarity has been proven in the previous Section \ref{sec:Jackpot-ergodicity}, devoted to data analyses.

Given the stationarity of the Jackpot time-series in the previous section,
we consider it ergodic for the remaining part of this study.
We assume that the Markovian process is characterized by a time-independent but possibly state-dependent stochastic transition matrix.

The successive values of the Jackpot values, $x(t)$, constitute the states of the
considered dynamical system. The distribution of the $x$ values at a time moment $t$
is described by the $\rho(x,t)$ density function. Following
the general considerations for the LGGR master-type equations~\cite{Biro-Neda,LGGR},
the time evolution of this distribution function writes as:
 \begin{equation}
 \frac{\partial \rho(x,t)}{\partial t}=-\frac{\partial}{\partial x} \left[ \mu(x) \rho(x,t) \right] - \gamma(x) \rho(x,t) +\langle \gamma(x) \rangle (t) \delta(x).
 \label{equation:master_gen}
 \end{equation}
Here $\mu(x)$ accounts for the state-dependent local growth rate
(probability per unit time to growth from state $x$ to a state
between $x$ and $x+dx$, for $dx=1$).
The reset is governed by the $\gamma(x)$ reset rate: by the probability that the system
resets from state $x$ to state $0$ in a unit time interval.
The last term of the evolution equation is the re-feeding
at $x=0$ imposed by the Dirac delta function $\delta (x)$.
This term serves to preserve the normalization of $\rho(x,t)$.
The mean value of the reset rate ($\langle \gamma \rangle$) is given as:
 \begin{equation}
 \langle \gamma(x) \rangle (t) = \int_{\{x\}}\limits\! \gamma(x) \rho(x,t) \, dx
 \label{equation:feeding}
 \end{equation}
The model is complete, it allows for a mathematical solution once the $\mu(x)$
and $\gamma(x)$ state-dependent rates are given.
The $\rho_s(x)$ stationary distribution is readily obtained by imposing
 $\partial \rho(x,t)/ \partial t=0$:
 \begin{equation}
\rho_s(x) \: = \: \frac{C}{\mu(x)} \, e^{-\int_{\{x\}}\limits\! \frac{\gamma(u)}{\mu(u)} \,du},
\label{stat-distr}
\end{equation}
with $C$ being a normalization constant. Convergence to the stationary state
$\rho_s(x)$ is proven for quite general conditions \cite{convergenceLGGR}.

In order to apply this model for the monitored Jackpot amounts, we need to obtain reset  $\gamma(x)$ and growth $\mu(x)$ rates realistically.
The increase per unit time (one week) in the Jackpot value is determined by the
number of sold tickets.
As the Jackpot value is higher, the number of sold tickets also rises, so one deals with a multiplicative growth.
The simple $\mu(x)=a(1+b \,x)$ linear form as a function of the $x$ Jackpot value
seems therefore to be a reasonable approximation.
The reset probability per unit time also depends on the number of sold tickets,
which in turn depends on the increase in the Jackpot values.
Taking into account that some of the sold tickets might not introduce new numbers
in the lottery,
for determining the $\gamma(y)$ reset probability for $y$ number of sold tickets
one should take into account the number of
differently completed lottery tickets (tickets with different numbers).
The probability of a reset (winning the jackpot) depends solely on the
number of differently completed lottery tickets. We thus write up a simple equation
\begin{equation}
\gamma(y+dy)=\gamma(y)+dy\, \alpha (1-\gamma(y)),
\label{gamma-evol}
\end{equation}
which states that by increasing the number of sold tickets with an amount of $dy$ the increase in the win-probability to reset the Jackpot depends on the $dy$ input amount and on the $1-\gamma(y)$ probability that each of them is not in the previous set of $y$. Equation (\ref{gamma-evol}) leads to the differential equation:
\begin{equation}
\frac{d \gamma(y)}{dy}=\alpha \left(1-\gamma(y) \right).
\end{equation}
Considering the imposed boundary condition $\gamma(0)=1-\beta$, we get the solution:
\begin{equation}
\gamma(y)=(1-\beta \, e^{-\alpha \,y})
\label{gamma_mu}
\end{equation}
Assuming that $y\propto \mu(x)$, we get the general form in which one should consider
the growth and reset rates:
\begin{equation}
\mu(x)=a(1+b \,x),
\label{growth-rate}
\end{equation}
 and
\begin{equation}
\gamma(x)= \left( 1-\beta e^{-\kappa (1+b\,x)} \right),
\label{reset-rate}
\end{equation}
with $\kappa \propto a \,  \alpha $.
Having these rates, it is straightforward to determine the $\rho_s(x)$ stationary
distribution. According to equation (\ref{stat-distr}) we obtain:
\begin{equation}
\rho_s(x) = C \, (1+b \,x)^{-\lambda-1} e^{\nu \,\text{Ei}\left(-\kappa(1+b\,x) \right)}
\label{distr}
\end{equation}
In the above form $\lambda=1/(a\,b)$, $\nu=\beta \lambda$, $\text{Ei}(x)$ denotes the
exponential integral function
\begin{equation}
\text{Ei}(x)=\int_{-x}^{\infty}\limits \!\frac{e^{-t}}{t} \, dt,
\end{equation}
and $C$ is a normalization constant, ensuring that $\int_{0}^{\infty} \rho_s(x) dx=1$.

%%%%%%%%%% HERE5
For the obtained results  some comments are now in order:
\begin{enumerate}
\item For $\beta>1$ the reset rate given by equation (\ref{reset-rate}) can be negative.
Indeed if
\begin{equation}
x<\frac{1}{b} -\frac{1}{\kappa \, b} \ln(\beta),
\end{equation}
the reset rate is negative. For an $x \ge 0$  value, this negative reset rate means
that after the reset process, the growth dynamics starts with a nonzero $x$ jackpot value.
The probability rate for such an event is $-\gamma(x)$.
As we will see later in the real-world data, this is the case for many lottery systems.

\item \textcolor{black}{The obtained probability density function has a maximum at
\begin{equation}
x_{max}=-\frac{\kappa + \ln(\frac{1+\lambda}{\nu})}{b\cdot \kappa}.
\end{equation}
When the dynamics is characterized by a "smart reset" process ($\beta > 1$ $\rightarrow$ $\nu > \lambda$), $\rho_s(x)$ can have a maximum on the $x \in (0, \infty)$ interval. In the following sections we show that this aspect is in agreement with the empirical observations. 
}

\item The asymptotic behavior ($x \rightarrow \infty$) of the $\rho_s(x)$ probability
density (equation (\ref{distr})) is governed by
the  $(1+b \,x)^{-\lambda-1}$ Tsallis-Pareto term,
since \textcolor{black}{$\lim_{x\rightarrow \infty}\text{Ei}(x)=0$}, and therefore
$\lim_{x\rightarrow \infty} e^{\nu \,\text{Ei}\left(-\kappa(1+b\,x) \right)}=1$.
This means that the tail of the distribution
of the Jackpot values has to scale with the exponent $-\lambda-1$.
The actual form of the power-law-like tail is determined solely by the $a$ and $b$
parameters governing the multiplicative growth.

\item The $C$ normalization constant cannot be written in an analytic form, and has to be computed numerically
after fixing the model parameters: $a$, $b$, $\kappa$, and $\beta$.
\end{enumerate}

\section{Discussions and validation of the elaborated model}
\label{sec:discus}

After having introduced the LGGR-model and the stationary jackpot time-series
for different lotteries, we argue for the applicability of our model.
%detailed in the previous section, for describing the experimental observations.
In order to make the understanding of the available Jackpot data more intuitive
and unified, we rescale each time series to the time average of the corresponding
Jackpot value, $z = \frac{x}{\langle x \rangle_{t}}$.

\begin{figure}[!ht]
    \centering
    \includegraphics[width=\textwidth]{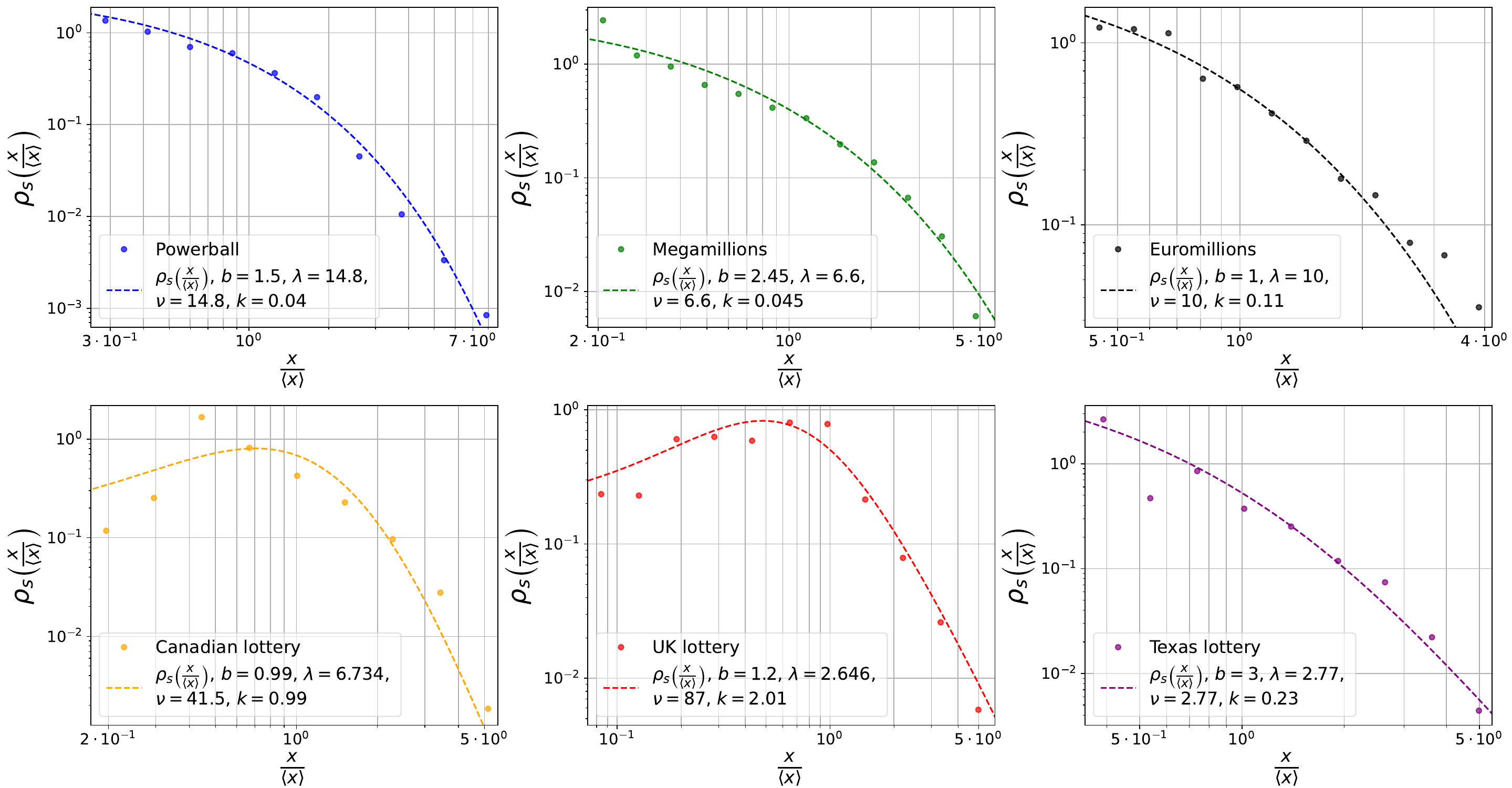}
    \caption{The stationary probability distribution of the Jackpot prizes $\rho_s (\frac{x}{\langle x \rangle_{t}})$ as a function of the mean rescaled
    Jackpot $\frac{x}{\langle x \rangle_{t}}$, for the six studied lotteries.}
    \label{fig5_dist}
\end{figure}

First, as a proof of concept, we verify whether the probability distribution function,
obtained as the stationary solution of the LGGR model (Equation~\ref{distr})
assuming preferential growth rate for the Jackpot values (Equation~\ref{growth-rate})
fits the \textcolor{black}{observed} probability distributions.
In Figure~\ref{fig5_dist}. the \textcolor{black}{empirical} distributions of the relative Jackpot values,
$\rho_s (z)$, are presented, along with the probability density function given by
Equation~\ref{distr}.
In the cases of the Canadian and the UK lotteries, the probability distribution's shape
differs considerably from the Jackpot distributions of the other four lotteries.
This qualitative distinction can be attributed to the distinct rule system.
Among others, the dissimilarity originates primarily from the different reset processes
characteristic of these lotteries.
While the Euro millions lottery and the lotteries played in the US are characterized by conventional, total reset processes to a minimal Jackpot value~\cite{Powerball_rules,Powerball_15_17,Powerball_18_22,Powerball_growth,Megamill_01_06,Megamill_06_12,Megamill_rules,texas_92_95,texas_96_98,texas_99_01,texas_rules} for the Canadian and UK lotteries a so-called "smart reset" process is typical~\cite{Can_lott,Uk_lotto}. \textcolor{black}{The different types of reset are recognizable also in the shape of the time series data presented in the A sections of Figure~\ref{fig2_lottery_time_series}.} In the cases when a total reset applies, the value of the $\beta$ parameter is $1$, leaving us with only three free parameters \textcolor{black}{ ($\nu=\lambda$). In this scenario $\rho_s(z)$ follows a monotonically decaying trend on the $x \in [0, \infty)$ interval. This monotonic decay of the distribution is also supported by the empirical probability distributions obtained for the four lotteries (Powerball, Megamillions, Euromillions, Texas lottery) characterized by conventional reset (Figure~\ref{fig5_dist}).}
The smart reset rate however, ($\beta \neq 1$) implies that after the Jackpot is won,
its value can be reset to values higher than the minimum Jackpot value. \textcolor{black}{This condition induces a peak of the probability density function in the $x \in (0, \infty)$ interval, as it is noticeable in the case of the Canadian and the UK lotteries.}
In the case of such lotteries, the smaller prizes, awarded for players matching
fewer numbers than the maximum possible lottery numbers,
are also paid from the Jackpot found~\cite{Can_lott, Uk_lotto}.

Alongside the difference in reset processes, the relatively large value of the
$w$ measure, defined in the introduction (see Table~\ref{table:lotteries}),
also suggests a deviation from the other lotteries.
This indicates that for these lotteries the total number of potential players is relatively high compared to the probability of winning the Jackpot, resulting in notably accelerated Jackpot dynamics.

%%%%%%%%% HERE7
The fitting parameters were chosen based on visual inspection, as the objective
here is not to provide a perfect fit for the \textcolor{black}{empirically} obtained probability
density functions.
This would be impractical due to the limited amount of \textcolor{black}{empirical} data available
and the fact that the rules of lotteries change from time to time,
while our modeling focuses on relatively short periods with unchanging conditions.
Rather, our objective is to construct a statistical mean-field model capable of
providing a consistent explanation for the dynamics governing the stationary
evolution of Jackpot values.

% a kappa = c*a ahol c >1 mert c = 1/(a jegyeknek aranak az a resze, ami noveli a jackpottot)

\subsubsection*{Preferencial growth rate and consistency of the
$\mathbf{a}$, $\mathbf{b}$ parameters}

The assumed form of the growth rate given in Equation~\ref{growth-rate}
contains two parameters that influence the form of the probability density function.
Considering the relation, $\lambda=1/(a\cdot b)$, between the $a$, $b$ and
$\lambda$ parameters, the value of the parameter $a$ can be deduced from the values
obtained from fitting the probability distributions of the relative Jackpot values
in Figure~\ref{fig5_dist}.

We comment here also on the meaning of the two parameters of the growth kernel function
from the point of view of the lottery. The parameters $a$ and $b$ are clearly characterizing
the player pool.
Parameter $a$ is linked to the size of the active player society, representing the proportion
of sales contributed by the subset of players who participate in the lottery regardless of
the Jackpot's value.
If we consider the form given by Equation~\ref{growth-rate} the value of $a$ can be given
as the extrapolated average sales after the lottery is won, $\mu(0) = a$.
The $b$ parameter on the other hand addresses the player attraction success, as in the linear growth rate (Equation~\ref{growth-rate}) the term $a\cdot b$ defines the slope according to which the sales increases with the Jackpot value.
As the Jackpot grows, sales also increase because the alluring prize encourages
the purchase of multiple tickets.

As mentioned in Section \ref{sec:Jackpot-ergodicity} the time-series contain all the information needed to elaborate the form of the
growth and reset rates.
Thus, we determined the form of the growth and reset rates based on the \textcolor{black}{real-world}
data and compare against the assumed kernel functions
(Equations~\ref{growth-rate} and~\ref{reset-rate}) with the chosen values of the
fitting parameters in Figure~\ref{fig5_dist}.

The growth rate was determined \textcolor{black}{empirically} by calculating the average relative
increase of the Jackpot values in between consecutive draws when no winners were announced.
Then, the calculated growth rate is represented in the function of the relative Jackpot
values averaged within consecutive relative jackpot value intervals.

\begin{equation}
\mu(\langle \frac{x}{\langle x \rangle_t} \rangle_z) = \langle \frac{1}{\langle x \rangle_t} \cdot \frac{\Delta x}{\Delta t}\rangle_z,
\label{growth_experiment}
\end{equation}
where $\langle \rangle_z$ denotes the averaging within bins defined based on the relative Jackpot value $z = \frac{x}{\langle x \rangle_{t}}$.

\begin{figure}[!ht]
    \centering
    \includegraphics[width=\textwidth]{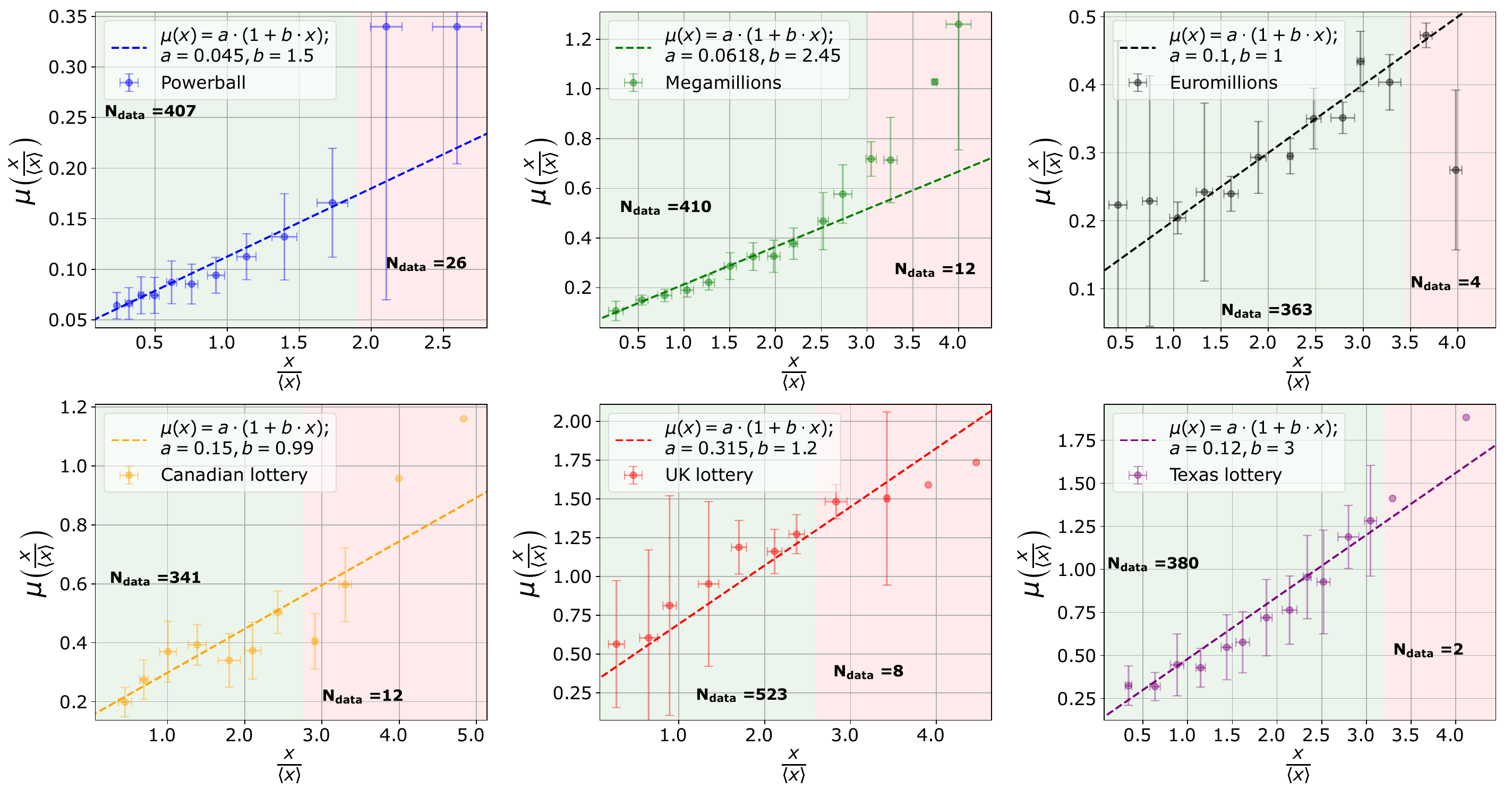}
    \caption{The \textcolor{black}{empirical} growth rate $\mu( x / \langle x \rangle_{t})$ presented as the function of $x / \langle x \rangle_{t}$, along with
the theoretical fit and the parameters for the studied lotteries.
The \textcolor{black}{empirical} data averages are presented along with the error bars for the
averages (standard deviation in the ensemble) in the bins for both axes.
We also present the number of data points, $N_{data}$, used for calculating the averages
in order to demonstrate that the red regions in the plots are not reliable for
statistical inference.}
    \label{fig3_growth}
\end{figure}

In Figure~\ref{fig3_growth} the \textcolor{black}{empirically} obtained growth rates, defined in
Equation~\ref{growth_experiment}, are represented.
On each figure, we indicated the uncertainty of each data point's positions using the
standard deviation of the data in the corresponding bins along both axis.
We split the graphs along the $x$ axis into two parts, based on the amount of data,
$N_{data}$, that fell into the respective regions.
By changing the background color of the figures, we indicate the reliability of the
data within the regions, green meaning reliable, while red meaning unreliable.
In each case, we fit parameters to the data only from the green region of the graphs.

The theoretical fitting curves shown in the subfigures of Figure~\ref{fig3_growth}
are given by Equation~\ref{growth-rate}, while for the values of the parameters
$a$ and $b$ we considered the ones used for fitting the probability distribution functions
in Figure~\ref{fig5_dist}.
\textcolor{black}{We highlight here that despite our linear assumption, in the case of some of the studied lotteries (Powerball, Megamillions, Texas lottery) the measured growth rates suggest a slightly superlinear trend (Figure~\ref{fig5_dist}). However, such a kernel function for the growth rate, that would better fit the empirical growth data, would make the analytical calculation of the stationary probability density function prohibitively tedious. A more complex form of the $\mu(x)$ function would make the integral in Equation~\ref{stat-distr} much more complicated due to the exponential relation between the growth and reset rates, defined by Equation~\ref{reset-rate}. Therefore we opted for the simpler linear approximation, which also shows a quite good alignment with the empirically determined growth rates, as depicted in Figure~\ref{fig3_growth}.}
%The assumed preferential growth rate shows a quite good alignment with the empirically determined growth rates, as depicted in Figure~\ref{fig5_dist}. This alignment supports the idea of linearly increasing growth rates in the context of lotteries. 
%The alignment between the assumed preferential growth rates determined using the fitting parameters in Figure~\ref{fig5_dist} and the \textcolor{black}{empirically} measured growth rates strongly supports the idea of linearly increasing growth rates in the context of lotteries.
Additionally, this finding validates the suitability and internal consistency of the proposed model.

\subsubsection*{Comments on the form of the assumed and \textcolor{black}{empirical} reset rates}
 
In Section \ref{sec:LGGR-general} we stated that there is a connection between the two rates that leads to the reset rate in the form of Equation~\ref{reset-rate}.
Fortunately, this connection between the rates resolves the difficulty of determining
the realistic form of the reset rate, as it is much more challenging to establish
\textcolor{black}{empirically} due to the infrequency of resetting compared to growth.
Moreover, the growth-reset relation introduces another two free model parameters,
$\alpha$ (or $\kappa = a\cdot \alpha$) and $\beta$.

The parameter $\beta$ dictates the character of the reset process.
When $\beta=1$, it results in a total reset, but if $\beta > e^{\kappa (1+b\cdot x_{c})}$,
then the reset becomes negative for $x<x_{c}$, evolving into a smart reset rate.
The negative reset rate indicates an incoming process into the system at the state
characterized by the value $x$.
The parameter \textcolor{black}{$\kappa$ is proportional to} the reciprocal of the unit price of the lottery tickets,
ensuring that in Equation~\ref{reset-rate}, the number of sold tickets appears in the exponent.

Due to the less frequent nature of resetting, especially in the case of lotteries
characterized by total reset and a small $w$ measure, there have not been enough
resetting events occurring during the studied time periods to construct and verify
the shape of the reset rate.

On the other hand, in the case of the Canadian and UK lotteries, due to their
accelerated dynamics (explained by a relatively large $w$) and the presence of a
smart reset allowing the Jackpot to be reset to values greater than the minimum Jackpot value
more frequently, we have a sufficient number of resetting records available to
construct the reset rate from real-world data. Another advantageous circumstance is that
these two lotteries offer the longest datasets with unchanging rulesets.

Because the connection between the reset rate and the growth rate (Equation~\ref{gamma_mu})
is inherent to the design of lottery games, it holds true for other types of lotteries as well.
As a result, we can confidently assume that the \textcolor{black}{empirically} determined shape of the
reset rate (provided we have sufficient data) could be matched with the theoretically
predicted kernel function for the reset rate for other lotteries, too.

\textcolor{black}{Empirically} the reset rate can be defined as a function of the scaled Jackpot value
$x/\langle x \rangle_t$.
It is computed by subtracting the total number of events when the Jackpot value is
reset from $x/\langle x \rangle_t$, from the total number of events when the Jackpot
is reset to the same value from a greater relative value.
The result is then divided by the total number of realizations when the relative
value of the Jackpot is $x/\langle x \rangle_t$.

\begin{equation}
\gamma(\frac{x}{\langle x \rangle_{t}}) = \frac{N_{win}(\frac{x}{\langle x \rangle_{t}})-N_{ret}(\frac{x}{\langle x \rangle_{t}})}{N(\frac{x}{\langle x \rangle_{t}})}
\label{eq:reset-exp}
\end{equation}

\begin{figure}[!ht]
    \centering
    \includegraphics[width=.8\textwidth]{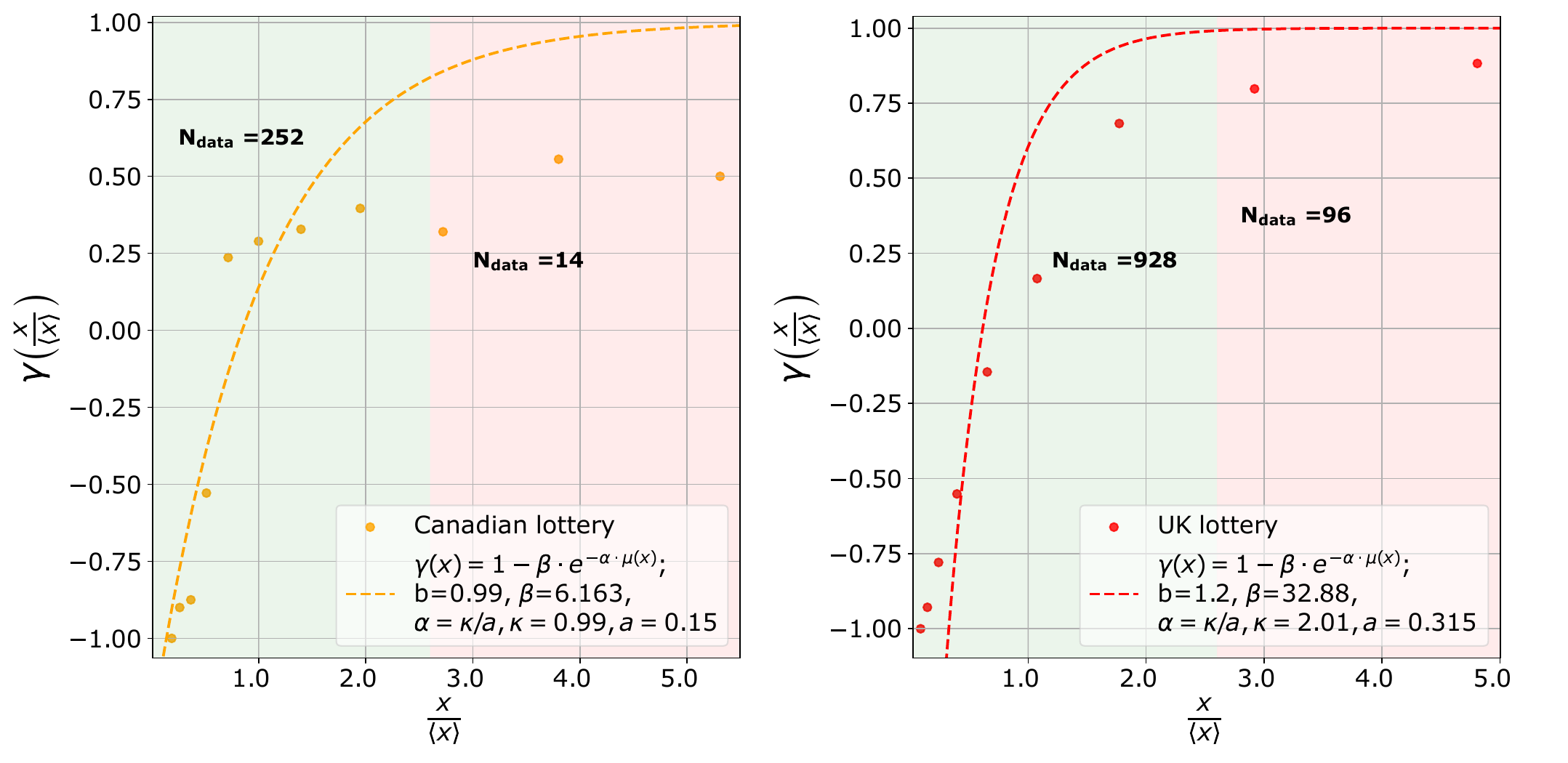}
    \caption{The \textcolor{black}{empirical} reset rate calculated by Equation~\ref{eq:reset-exp},
	fitted with the theoretical reset rate,
        for the Canadian lottery and UK lottery.
	We used the parameters obtained for fitting the probability density functions.}
    \label{fig4_reset}
\end{figure}

In Figure~\ref{fig4_reset} the \textcolor{black}{empirically} determined reset rates are shown
for the Canadian and UK lotteries, as a function the mean rescaled Jackpot
$x / \langle x \rangle_{t}$.
Once more it's important to highlight that the last few bins in these plots have a
small number of data points, making this part of the plot less reliable. 
%To demonstrate the consistency of the general assumptions of the model and the chosen parameter values, similar to the case of the growth rate, we fit the experimental reset rates with the reset rate considered to design the model.
As we did with the growth rate, we fit the \textcolor{black}{empirical} reset rates using the reset rate
form incorporated in the model's design to demonstrate the consistency of the model's
general assumptions regarding the reset and the chosen parameter values.

%The theoretical fit plotted with the experimental data is given by Equation~\ref{reset-rate}, here again using the parameters considered for fitting the probability distribution function of the relative Jackpot values in Figure~\ref{fig5_dist} and remaining with a %single free parameter, $\alpha$.

The theoretical fit, displayed alongside the \textcolor{black}{empirical} data, is governed by
Equation~\ref{reset-rate}, using the same parameters employed for fitting the
probability distribution function of the relative Jackpot values in Figure~\ref{fig5_dist}
and retaining only one free parameter, $\alpha$.
%Although the matching of the curves is \textcolor{black}{good}, it is a remarkable result given the poor quantity of available data.
% \textcolor{black}{The matching of the curves is good, it is a remarkable result despite the poor quantity of available data.}
 \textcolor{black}{The matching of the curves is good, it is a remarkable result despite the poor quantity of available data. Furthermore, we assumed that the lottery tickets are completed independently from each other. In reality, however, there might be correlations in the completed combinations, which would further diminish the rate of reset. This is in agreement with the results represented in Figure~\ref{fig4_reset}, where the empirical reset rate shows a somewhat slower increasing trend than the theoretical one. Another cause of the poor alignment might also be the imperfect fitting of the growth rates with the proposed linear function (Equation~\ref{growth-rate}). As Figure~\ref{fig3_growth} illustrates, the fitting of the empirical growth rates for both Canadian and UK lotteries does not display perfect agreement, and this discrepancy may be amplified when fitting the empirical reset data in Figure~\ref{fig4_reset}.
} 
%As Figure~\ref{growth-rate} shows the fitting of the empirical growth rates in the case of the Canadian and UK lotteries does not show a perfect agreement either, and this discrepancy may be amplified when fitting the empirical reset data in Figure~\ref{reset-rate}.
%Besides the poor data quality, a possible explanation for the poor alignment between the data and the predicted reset rate is that when we derived the form of the reset rate (Equation~\ref{gamma_mu}),

\section{Conclusions}
\label{sec:conclusion}

The value of the Jackpot in lotteries is constantly followed by millions of people,
yet there are very few of them who think of it as a complex system.
Although the format of the lotteries can differ, the resulting dynamics of the
combined system of players and the ruleset defined by the lottery associations is very similar.
%A simple measure that might suggest the similarity of lotteries is the  $w = P_w \times Pop.$, the value of which does not vari much from lottery to lottery, by this keeping the winning frequency at an optimum.
The simple rules, combined with the behavior of the players, result in an emergent,
non-deterministic evolution of the Jackpot prize. The dynamics of the Jackpot is
governed by a persistent growth process, that is the result of the ticket sales and
a probabilistic reset rate, induced by the rare winnings.
This stochastic dynamics serves as a classic example of phenomena that can be
effectively described using the LGGR model.
In the case of the lottery, however, instead of having a system with multiple elements
that are acting at the same conditions (considering a mean-field type model),
there is only a single component behaving according to the same rules.
The dynamics of this single element is entirely characterized by the Jackpot time-series.

In order to be handled as an ergodic Markovian process, we considered such fractions of
the entire historical data, for that the rules and the size of the player pool were
stationary  (not changing the rules and maintaining the $w$ value).
For such time periods the lottery can be considered ergodic as we have demonstrated by proving their stationarity and aperiodic nature (Figure~\ref{fig2_lottery_time_series}).
The design of the LGGR model required the establishment of both the growth and reset rates.
As we have demonstrated in the previous sections, it is an inherent property of the
systems featuring the connection $\gamma(y)=(1-\beta \, e^{-\alpha \cdot \mu(y)})$
between the growth and reset rates.
Being aware of this, it is sufficient to identify the form of the growth rate only.

For choosing a linear growth rate, our intuition was tested and proved to be correct
by real-world data (see Figure~\ref{fig3_growth}).
The relation stated in Equation~\ref{gamma_mu} between growth and reset rates was also
validated by the \textcolor{black}{observed} reset rate, cf. Figure~\ref{fig4_reset}.
\textcolor{black}{In the discussion we have commented on the fitting quality of the empirical growth and reset. In the case of the growth rate, although the data might show a faster increasing trend (e.g., Megamillions lottery), the proposed linear function provides a good fit and facilitates the analytical calculation of the stationary probability density function. The growth and reset rates are connected through Equation~\ref{reset-rate}. For the reset rate, the imperfect alignment in the fitting of the growth rate results in an inherent difference between the empirical data and the proposed reset kernel function. Our model works well even with this limited amount of data.}

By using these rates, both proven to be realistic by \textcolor{black}{empirical} data (Figure~\ref{fig3_growth} and Figure~\ref{fig4_reset}), in the stationary limit, the LGGR model produced a probability distribution function (Equation~\ref{distr}) that with the right parameters fits in a totally consistent manner,
the observed distributions of the Jackpot values in all six lotteries we studied.
A notable feature of this model is its ability to depict two types of lottery dynamics:
one defined by a total reset (Euromillions, Powerball, Megamillions, Texas lottery) and
the other by an incomplete reset (UK Lotto and Canadian Lotto) process.
\textcolor{black}{In the case of the different lotteries, characterized by the same reset scenario, the dynamics seems to be similar, resulting in similar shaped probability density functions (Figure~\ref{fig5_dist}). Within a reset scenario, the shape of the $\rho_s(x)$ stationary distribution is qualitatively stable regarding the parameter values. Although the $\beta$ parameter clearly captures a lottery rule characterizing the reset dynamics, which has a direct influence also on the shape of the stationary distribution, the connection between the values of the other parameters and the lottery rules is not trivial. For a better understanding of the connection between the parameter values and the actual lottery rules, a more comprehensive study of the specific lotteries would be needed.}
%The type of the reset depends on the $\beta$ parameter introduced through Equation~\ref(reset-rate). 

Another observation is worth noting here, related to the straightforward $w$ measure,
which despite its apparent simplicity can effectively capture some key properties of lotteries.

As shown in Table~\ref{table:lotteries}, the $w$ value is consistently maintained at an
optimal low value ($0.5<w<3$), determining the ideal speed of the dynamics,
and keeping the players interested.

\section*{Acknowledgements}

This work was supported by the project
“A better understanding of socio-economic systems using quantitative methods from Physics”
funded by the European Union – NextgenerationEU and the Romanian Government, under
National Recovery and Resilience Plan for Romania, contract no 760034/23.05.2023,
code PNRR-C9-I8-CF255/29.11.2022, through the Romanian Ministry of Research, Innovation
and Digitalization, within Component 9, Investment I8. The work of Sz.K.  is also supported by
the Collegium Talentum Program of Hungary.

\section*{Author contributions statement}
T.S.B. and Z.N. have conceptualized and created the LGGR model. Its application to lottery
Jackpot time series was suggested by I.G. Data curation and analysis were carried out
by I.G. and Sz.K., supervised by Z.N. Visualization by I.G. and Sz.K., The manuscript has been written by Sz.K. and I.G.
and Z.N., edited and corrected by T.S.B. All authors contributed to discussing the results
and form of presentation.

\section*{Competing interests}

The authors declare no competing interests.

\section*{Data availability}
The data used in this study are available at the referenced internet links.
The data curated by us used in this study can be found in the following online repository: \url{https://doi.org/10.6084/m9.figshare.24310669}.

%\section*{References}

\end{document}